\documentclass[onecolumn. oneside. notitlepage]{article} 
\usepackage[english]{babel}
\usepackage[utf8]{inputenc}

\usepackage[gen]{eurosym}	%Daje możliwość użycia \euro
\usepackage{amsfonts}
\usepackage{graphicx}
\usepackage{url}
\usepackage{textcomp}
\usepackage{enumerate}
\usepackage{nccmath} 
\usepackage{xspace}
\usepackage{float}

\makeatletter \renewcommand{\thetable}{\@Roman\c@table}

\newcommand{\bea}{\begin{array}}
\newcommand{\eea}{\end{array}}
\newcommand{\be}{\begin{equation}}
\newcommand{\ee}{\end{equation}}
\newcommand{\ba}{\begin{eqnarray}}
\newcommand{\ea}{\end{eqnarray}}
\newcommand{\baw}{\begin{eqnarray*}}
\newcommand{\eaw}{\end{eqnarray*}}
\newcommand{\bt}{\begin{tabular}}
\newcommand{\et}{\end{tabular}}
\newcommand{\bc}{\begin{center}}
\newcommand{\ec}{\end{center}}
\newcommand{\ben}{\begin{enumerate}}
\newcommand{\een}{\end{enumerate}}
\newcommand{\bi}{\begin{itemize}}
\newcommand{\ei}{\end{itemize}}
\newcommand{\bmpage}{\begin{minipage}}
\newcommand{\empage}{\end{minipage}}

\newcommand{\brak}{\langle}
\newcommand{\kket}{\rangle}

\def \Id {{\rm Id}\,}

\def \Del{\Delta}

\def\sgr{{\bf s} }
\def\0gr{{\bf 0} }

\def\Sgr{{\bf S} }

\def\Hcal{ {\cal H} }

\title{Inequalities between ground-state energies of Heisenberg models} 

\begin{document} 
\renewcommand*{\figurename}{Figure}
\renewcommand*{\tablename}{Table}

%v2: Do czasu wyjasnienia, sa rozwazane tylko te same geometrie
% a rozne wielkosci ukladow

\author{{\em Jacek Wojtkiewicz} and {\em Rafał  Skolasiński${^1}$}
\\ \\
Department for Mathematical Methods in Physics\\
 Faculty of Physics, University of Warsaw \\
 Ho\.{z}a 74, 00-682 Warszawa, Poland\\
 e-mail: wjacek@fuw.edu.pl 
\\\\
${^1}$\footnotesize Current adress: Kavli Institute of NanoScience, 
Delft University of Technology,\\
\footnotesize Lorentzweg 1, 2628 CJ Delft, The Netherlands, 
e-mail: r.j.skolasinski-1@tudelft.nl
}

 \maketitle 
\abstract{

The Lieb-Schupp inequality is the inequality between ground state energies of certain antiferromagnetic Heisenberg spin systems. In our paper, the numerical value of energy difference given by Lieb-Schupp inequality has been tested for spin systems in various geometries: chains, ladders and quasi-two-dimensional lattices. It turned out that this energy difference  was strongly dependent on the class of the system. The relation between this difference and a fall-off  of a correlation function has been empirically found and formulated as a conjecture.
}%EndOfAbstract

%%%%%%%%
\section{Introduction}
\label{sec:Introd}
%%%%%%%%

The list of general results on the area of quantum spin systems is not too large. Among them,
there is a result due to Schupp \cite{Schupp}, establishing the inequality between a
ground-state energies of an antiferromagnetic Heisenberg chains. (The simplest example is 
the difference $2E_{m+n}$ and $E_{2n}+E_{2m}$, where $E_k$ is the ground state energy of
the Heisenberg chain with $k$ sites). Later on, this inequality has been 
extended to more general class of Heisenberg models \cite{LSW}. 

These inequalities are rigorous ones. They are based on the matrix inequality proved by Kennedy,
Lieb and Shastry (KLS) \cite{KLS2}. Authors applied this inequality to proving the existence of the Long Range Order in a class of anisotropic Reflection-Positive $d\geq 2$ quantum Heisenberg
models. The KLS inequality has also been applied to establish certain properties of ground states of
Hubbard models \cite{LiebHuM,SQT,Shen} as well as to prove the absence of
orderings in Heisenberg models on pyrochlore lattices \cite{LiebSchupp1,LiebSchupp2}.

Although the Schupp inequalities between ground-state energies of Heisenberg models are rigorous, they do not give any information about the actual value of this difference. It would be very intersting to test how close to each other are both sides of the inequality. This was one of the goals of our paper: {\em To test a numerical value of the difference between the both sides of the inequality.}

We have tested this difference numerically. We used exact diagonalization procedure and 
in some cases the DMRG method. The ARPACK++  and ALPS packages have been used~\cite{ARPACKref, ALPSref}.
We considered systems  in various geometries: chains, ladders, and rectangles, up to 27~-~28 sites\footnote{Depending on the lattice type.}  
(exact diagonalization) and 200~-~256 sites (DMRG). 
As a rule, the differences between both sides of inequality were {\em very small}. More quantitative
considerations allowed us to pose certain conjecture between strength of spin correlations
in the system and the difference between both sides of inequality. 
%We have also noticed some implications
%of the Schupp inequality for the properties of analogon of the {\em solvation force} in the quantum
%spin systems. 

The outline of the paper is as follows.

In the Sec.~\ref{sec:Formulations}, the ground-state energy inequalities have been
formulated.

In the Sec.~\ref{sec:Dokladnosc}, the 
differences between both sides of inequality for various systems have been numerically tested.
The chains, ladders, and rectangles in various geometries: square, pyrochlore and squares with crossing bonds,
have been analysed. The results and their implications (both rigorous and conjectural)
 have been presented.  

The Sec.~\ref{sec:Summary} contains summary, conclusions as well as perspectives for future
research.

%%%%%%%%
\section{Formulation of inequalities}
\label{sec:Formulations}
%%%%%%%%
%%%%%%%%%%%
\subsection{Structure of the system}
%%%%%%%%%%%
We consider finite lattice spin systems  with all  spins being finite, 
 so all spaces are finite-dimensional and operator are matrices. We assume that
 all spins are identical (this assumption can be relaxed).

We consider system which can be divided into two subsystems: $L$
('left') and $R$ ('right') parts. Corresponding division of the total Hilbert space $\Hcal$
is:
\be
\Hcal = \Hcal_L\otimes \Hcal_R,
\label{HilbertSp}
\ee
where $\Hcal_L$ and $\Hcal_R$ are Hilbert spaces for the $L$ and $R$ subsystems, respectively.
Let $H_L$ be the Hamiltonian of the $L$  subsystem acting on $\Hcal_L$,
and analogously $H_R$  the Hamiltonian of the $R$  subsystem acting on $\Hcal_R$.
They can be lifted to operators acting on the whole $\Hcal$:
\be
H_L \to H_L\otimes \Id_R,
\;\;\;\;\;
H_R\to \Id_L\otimes H_R.
\label{HLHR}
\ee
 Then, the Hamiltonian is of the following general form:
\be
H=H_L\otimes \Id_R +\Id_L\otimes H_R +H_I,
\label{GenHam}
\ee
where $H_I$ is the 'interaction' Hamiltonian, acting on $\Hcal$.
%We hope that it will be not confusing to denote $H_L$ and $H_L\otimes \Id_R$ by one common
%symbol $H_L$ (and analogously for $H_R$).

%%%%%%%%%%%
\subsection{Assumptions on Hamiltonian}
%%%%%%%%%%%
Hamiltonians $H_L$ and $H_R$ are arbitrary spin-interaction Hamiltonians.
The crucial assumption is the one concerning
the interaction Hamiltonian $H_I$. 

We assume \cite{Schupp} that it is the sum of {\em antiferromagnetic
Heisenberg interactions}. More precisely:

Let $A\subset L$ be some $m-$site subset of  $L$:
 $m=|A|$. Let us write: $A=\{i_1,i_2,\dots,i_m\}$. 
Let us define $\Sgr_A$ being operator acting on $\Hcal_L$:
 $\Sgr_A =\sum_{k=1}^m J_k \sgr_{i_k}$,
where coefficients  $J_k$ are {\em real}, and $\sgr$ is total spin operator:
$\sgr = (s^x,s^y,s^z)$.  Let $A'\subset R$ be some subset of $R$,
let $m=|A'|$, let $A'=\{i'_1,i'_2,\dots,i'_m\}$,
and define $\Sgr_{A'}$ -- an operator acting on $\Hcal_R$:
 $\Sgr_{A'} =\sum_{k=1}^m J_k \sgr_{i'_k}$. (So, we can say that
 $\Sgr_{A'}$ is a 'mirror image' of $\Sgr_{A}$; but we {\em do not} assume that the $L$
 system is a mirror image of $R$ system).
  Then we assume that the interaction 
 Hamiltonian is of the form
\be
H_I =\sum_A \Sgr_A\cdot \Sgr_{A'},
\label{InteractionHam}
\ee

%\vskip.5cm
%{\bf RYS. 1 : Ilustracja hamiltonian/ow oddzia/lywania
% (ulepszony rys. 1.7 ze Schuppa).}
%\vskip.5cm

\begin{figure}[!ht]
\center
\includegraphics[width=1.0\textwidth]{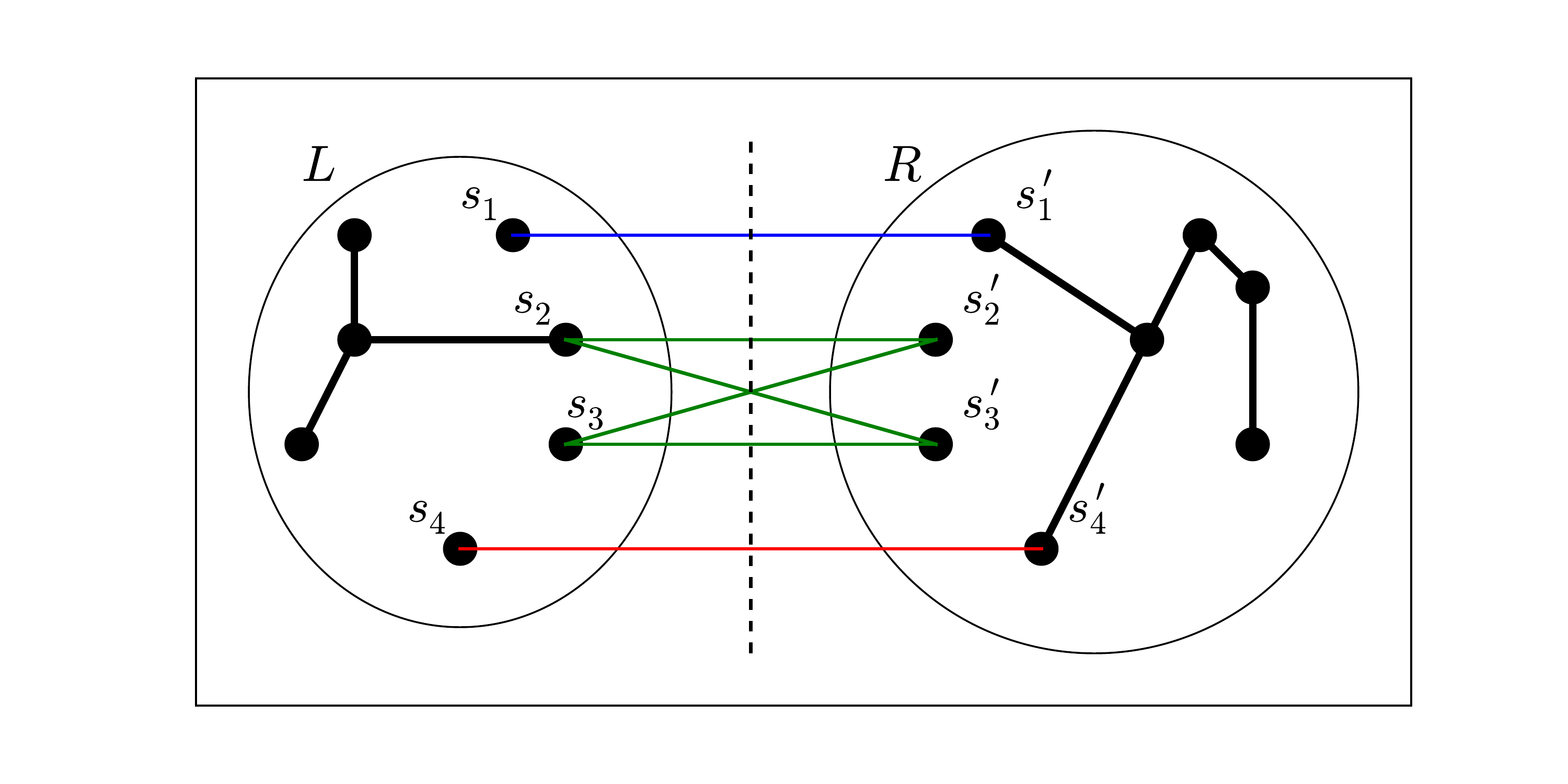}
\caption{Illustration of the structure of the interaction Hamiltonian.
Interactions within $L$ and $R$ subsystems are arbitrary,
but the interactions between them are antiferromagnetic, i.e. they have form
(\ref{InteractionHam}).}
\label{rys:StrukturaOddzialywan}
\end{figure}

The simplest example of $A$ is one site; it leads to ordinary Heisenberg AF interaction.
For $A$ being two-site set, an example is the pyrochlore lattice.
We will consider these two sorts of interactions.

%%%%%%%%%%%
\subsection{Formulation of the main inequality}
%%%%%%%%%%%
Consider now the following Hilbert spaces and systems ('$LL$' and '$RR$' ones):
\be
\Hcal_{LL} = \Hcal_L\otimes \Hcal_L,
\;\;\;\;\;
\Hcal_{RR} = \Hcal_R\otimes \Hcal_R;
\ee
and Hamiltonians:
\be
H_{LL}=H_L\otimes \Id_L + \Id_L\otimes H_L +\sum_A \Sgr_A\cdot \Sgr_{A},
\label{H_LL}
\ee
\be
H_{RR}=H_R\otimes \Id_R + \Id_R\otimes H_R +\sum_{A'} \Sgr_{A'}\cdot \Sgr_{A'}
\label{H_RR}
\ee
Let $E_{LL}$ be the ground-state energy of the Hamiltonian (\ref{H_LL}), 
$E_{RR}$ -- the ground-state energy of (\ref{H_RR}) and $E_{LR}$ -- the ground-state energy
of (\ref{GenHam}) with (\ref{InteractionHam}).

Under assumptions above, it has been proved in \cite{Schupp} that the following
inequality between ground-state energies $E_{LL}$, $E_{RR}$, $E_{LR}$ holds:
\be
E_{LL}+E_{RR}\leq 2 E_{LR}.
\label{MainIneq}
\ee

%\vskip.1cm
%{\bf RYS. 2 -- ilustracja podstawowej nierownosci (\ref{MainIneq}).}
%\vskip.1cm

\begin{figure}[!ht]
\center
\includegraphics[width=1.0\textwidth]{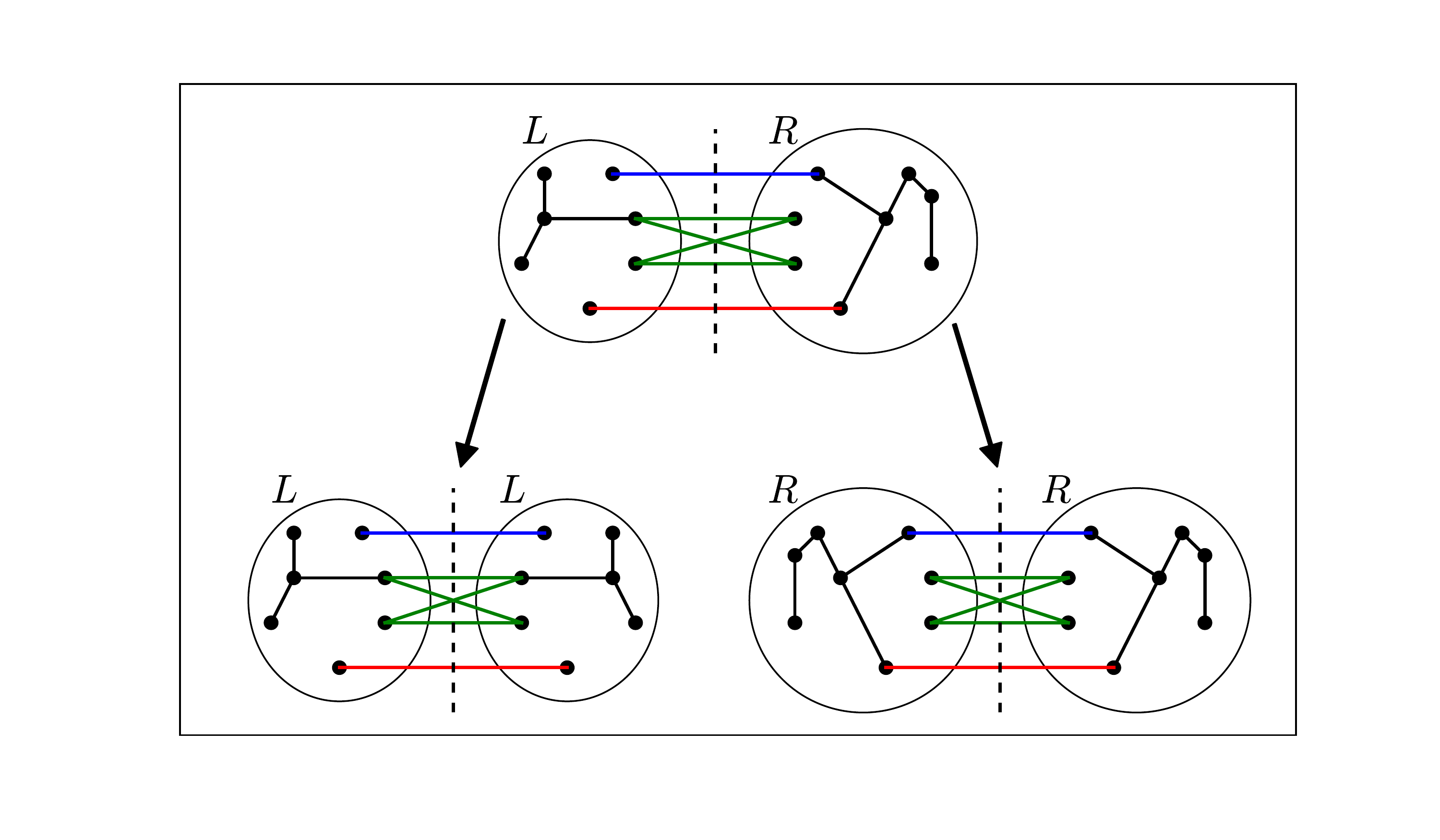}
\caption{Illustration of interrelations between initial $L-R$ system
and resulting $L-L$ and $R-R$ systems. In all three cases structures of 
the interaction
Hamiltonians are identical.}
\label{rys:OgolnyPodzial}
\end{figure}
%%%%%%%%

\section{Magnitudes of differences between two sides of inequality}
\label{sec:Dokladnosc}
%%%%%%%%
In this Section we present results for one-half spin Heisenberg models in chosen 
geometries. We test numerically how close to each other are the two sides of inequality~(\ref{MainIneq}).

We have examined the following systems: chains, ladders (square, pyrochlore and with
crossing bonds) as well as rectangles.

The main numerical method we have applied are: ED (Exact Eiagonalization)
as implemented in the  ARPACK++ package,\footnote{For some small system (up to 16 sites) ED was also performed with the ALPS library for a test purpose.} based on Lanczos algorithm with Arnoldi modification~\cite{ARPACKref}.
We present results of exact diagonalization for systems up to 28 sites. The precision of
the method is close to numeric limitation, which is~$10^{-14}$.

Moreover, for chosen class of systems (chains and ladders) we have also applied
the DMRG method as developed in the  ALPS package. We considered systems up to 256 sites. 
However, the price paid was some lack of precision, but still it was about~$10^{-10}$.

We have checked that in {\em all} tested systems the inequality was fulfilled in the
range of precision of the method (it must be so because the inequality is rigorous).
However it turns out that the magnitude of difference between two sides of inequality
strongly depend of the kind of 
the system under consideration. Below we present results we have obtained in more details.

%%%%%%%%%%%%%%%%%%%%%%%%%%%%%%%%%%%%%%%%%%%%%%%%%%%%%%%%%%%%%%%%%%%%%%%%%%%%%%%%%%%%%%%
\subsection{Chains}
%%%%%%%%%%%%%%%%%%%%%%%%%%%%%%%%%%%%%%%%%%%%%%%%%%%%%%%%%%%%%%%%%%%%%%%%%%%%%%%%%%%%%%%

In this subsection we present results of the analysis done for open chains. The coupling 
constant is equal to 1.

The results of exact diagonalization for chains of length up to $28$ sites are
presented in Table~\ref{tab:gsChains} in Appendix A. They are fully consistent with existing data for smaller chains
(see for instance \cite{Caspers}).
We  have also performed DMRG calculations for chains up to 200 sites.
We have checked correctness of DMRG calculations by comparison with ED results wherever it was possible and they were identical within numerical precision.

\begin{figure}[!ht]
\center
\includegraphics[width=1.0\textwidth]{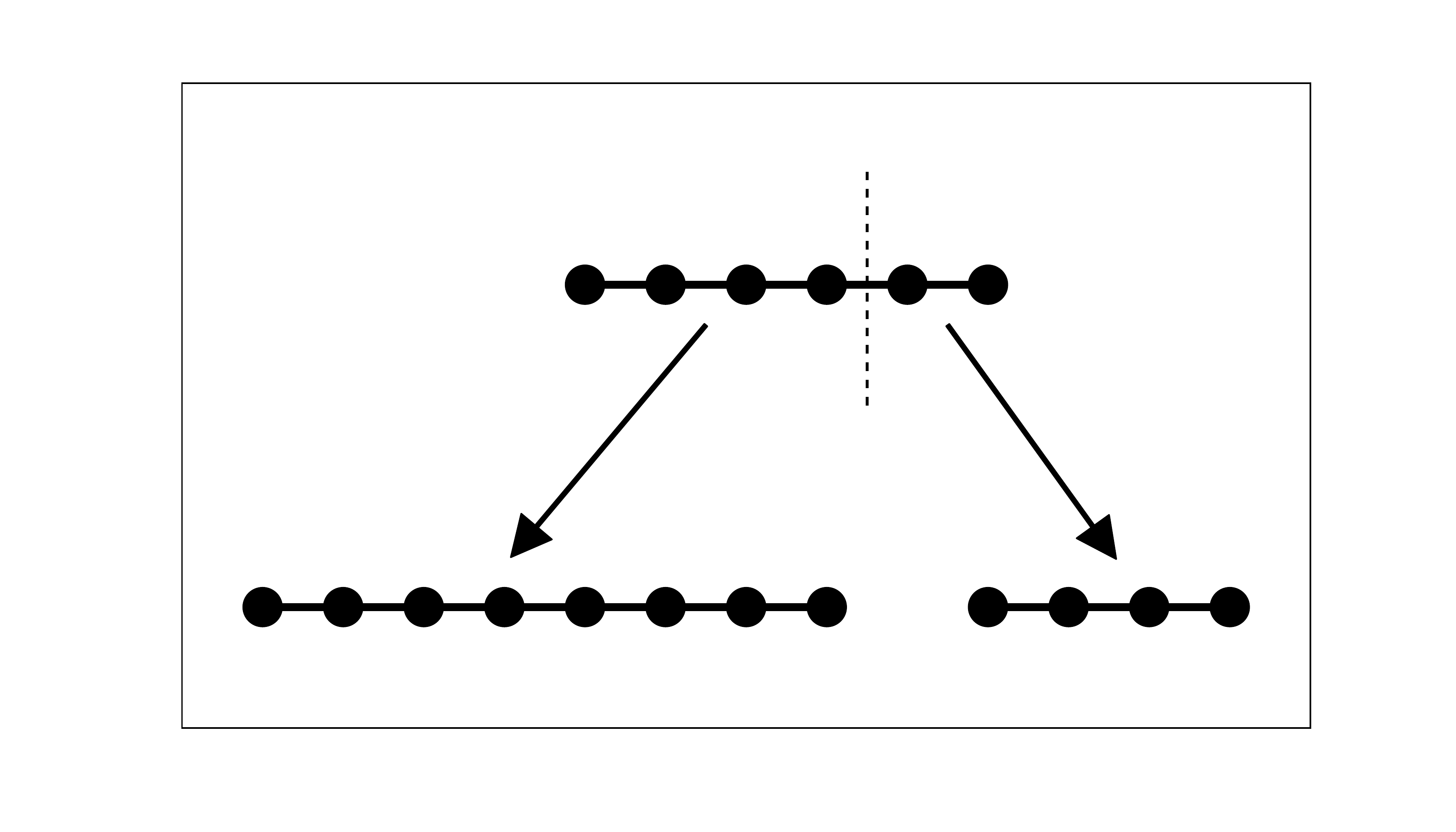}
\caption{Scheme of the open chain lattice and demonstration of the division. Chain of the 6 sites is divided into chains with 4 and 2 sites which further are 'glued' together
in combinations: 4+4 and 2+2, so 
 two chains of length of 8 and 4 sites are formed.}\label{rys:SchemeChain}
\end{figure}

To compare both sides of the Schupp inequality (\ref{MainIneq}), let us denote by $E_k$
the ground-state energy for the $k-$site chain. Then the inequality (\ref{MainIneq})
takes the form: (see Fig.~\ref{rys:SchemeChain} for illustration)
\be
2E_{m+n} \geq E_{2n}+E_{2m}.
\label{LSchainsI}
\ee

To better illustrate the results, let us rewrite the inequality (\ref{LSchainsI}) into the following form
\begin{equation}
\Delta_{mn} = 2E_{m+n} - E_{2n} - E_{2m}\geq 0,
\label{LSchainsII}
\end{equation}
and then slightly redefine $\Delta_{mn}$ into 
\begin{equation}
\Delta_{mn}\equiv\Delta^{(d)}_L = 2E_{L}-E_{L-2d}-E_{L+2d}\geq 0,
\label{LSchainsIII}
\end{equation}
where $L=m+n$ is the length of the initial chain, and $d$ is a distance from the middle of the chain where we are making a slice. For example: in Fig.~\ref{rys:SchemeChain} we have $L=6$ and $d=1$.

\begin{figure}[!ht]
\center
\includegraphics[width=1.0\textwidth]{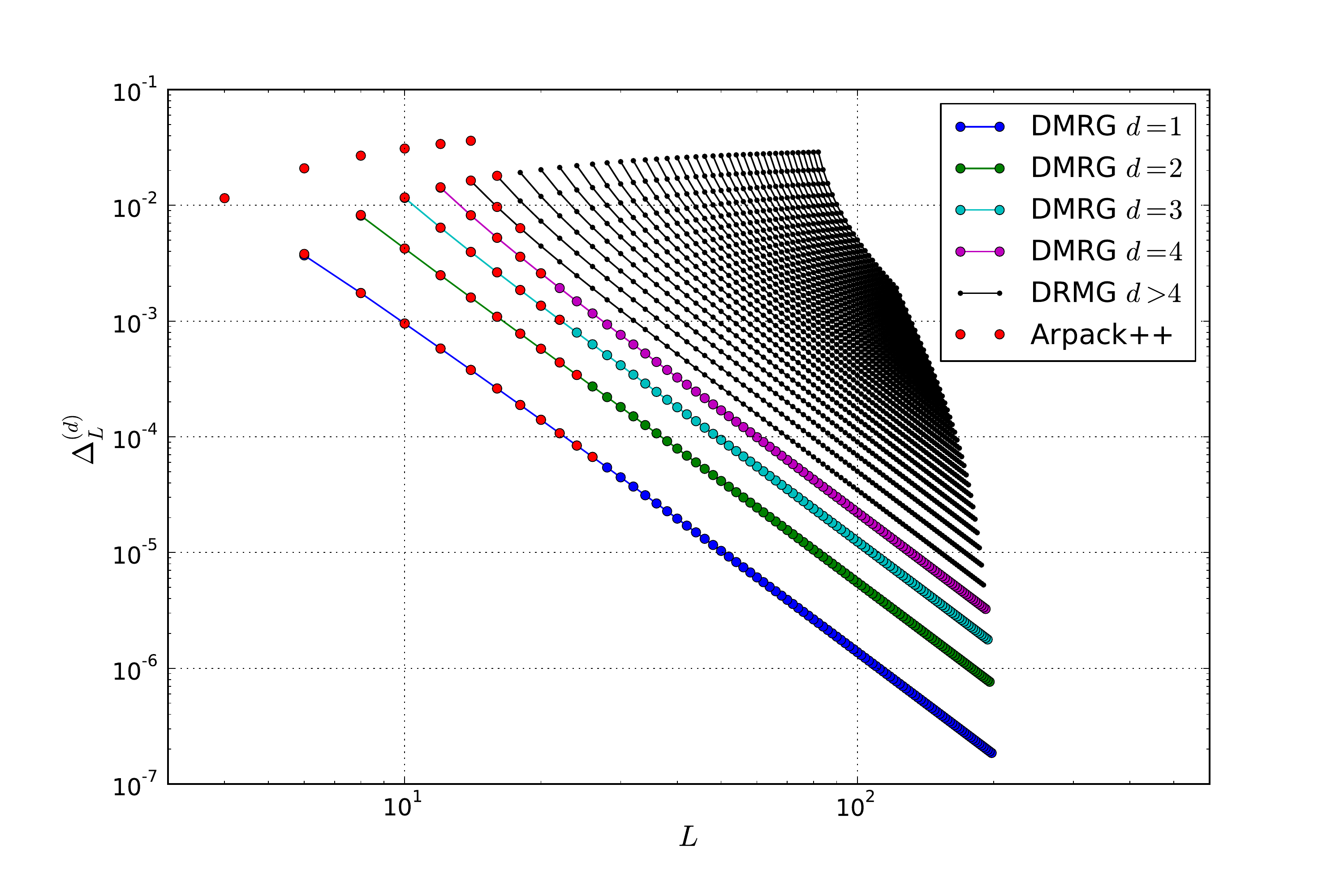}
\caption{ Open chains: plot of $\Delta^{(d)}_L$ for as a function of $L$ for different values of $d$. 
The scales on both axes are logarithmic.
The illustrated data have been obtained
 by exact diagonalization  and DMRG. %The closer to zero it is the better.
 }\label{rys:res_chains}
\end{figure}

In Fig.~\ref{rys:res_chains} we are making a plot of LHS of the inequality~(\ref{LSchainsIII}). It is clearly seen that the larger systems are, the difference $\Delta^{(d)}_L$ is smaller. Analogously, the more similar left and right subsystems are, the difference $\Delta^{(d)}_L$ 
is smaller too. The the difference $\Delta^{(d)}_L$ ranges from $10^{-2}$ (for few-site systems)
to  $10^{-6}$ (for hundred site systems divided about the middle).

One can ask how the differences $\Delta^{(d)}$ depend of the length $L$. One can guess that this
dependence can be {\em power law} or {\em exponential}. It turns out the first possibility is realized 
with good accuracy (see Fig.~\ref{rys:res_chains}). Assuming power-law dependence:
\be
\Delta^{(d)}_L \sim L^{-\theta}
\label{DeltaLchain}
\ee
we have found by least-square fitting that for $d=1,2,3,4$ the value of
$\theta=2.83(4)$ 

Let us make also the following comment about content of the Fig.~\ref{rys:res_chains}. Presented points refer to the situation where  lengths of initial chains are {\em even}. (Also all other systems that will be presented later are refered to lattices of even length.) We have also examined
 chains of the odd length. In this case, the data obtained by the exact diagonalization methods
  show that values of $\Delta^{(d)}_L$ are also greater than zero and have
   magnitude around $10^{-1}$. This behaviour is the same as claimed by the folk knowledge
on the area of antiferromagnetic spin systems: For even number of spins, they exhibit tendence to 
be paired, so in general the average energy per spin is lower for systems with even number of spins
than for systems with odd number. We confirm this rule for chains. 

{\em Remark.} One can ask whether the inequality (\ref{LSchainsI}) inequality can
be extended to the inequality
\be
2E_{(m+n)\slash{}2} \geq E_{n}+E_{m},
\label{LSfake}
\ee
where $m$ and $n$ are odd numbers. The answer is {\em no}. For instance, we have (see Table~\ref{tab:gsChains}): $2E_6=-4.97$, $E_5+E_7=-4.76$, so the inequality (\ref{LSfake}) cannot
be fulfilled in general.

%%%%%%%%%%%
\subsection{Square ladders}
%%%%%%%%%%%
The results of this subsection are based on exact diagonalization for ladders
of length up to 14 (i.e. 28--sites systems) and DMRG results for ladders of length up to 126 (i.e. 252--sites systems).
All coupling constants are equal to 1.
The ED results are collected in Table~\ref{tab:quasi2d} in Appendix~A.

\begin{figure}[!ht]
\center
\includegraphics[width=1.0\textwidth]{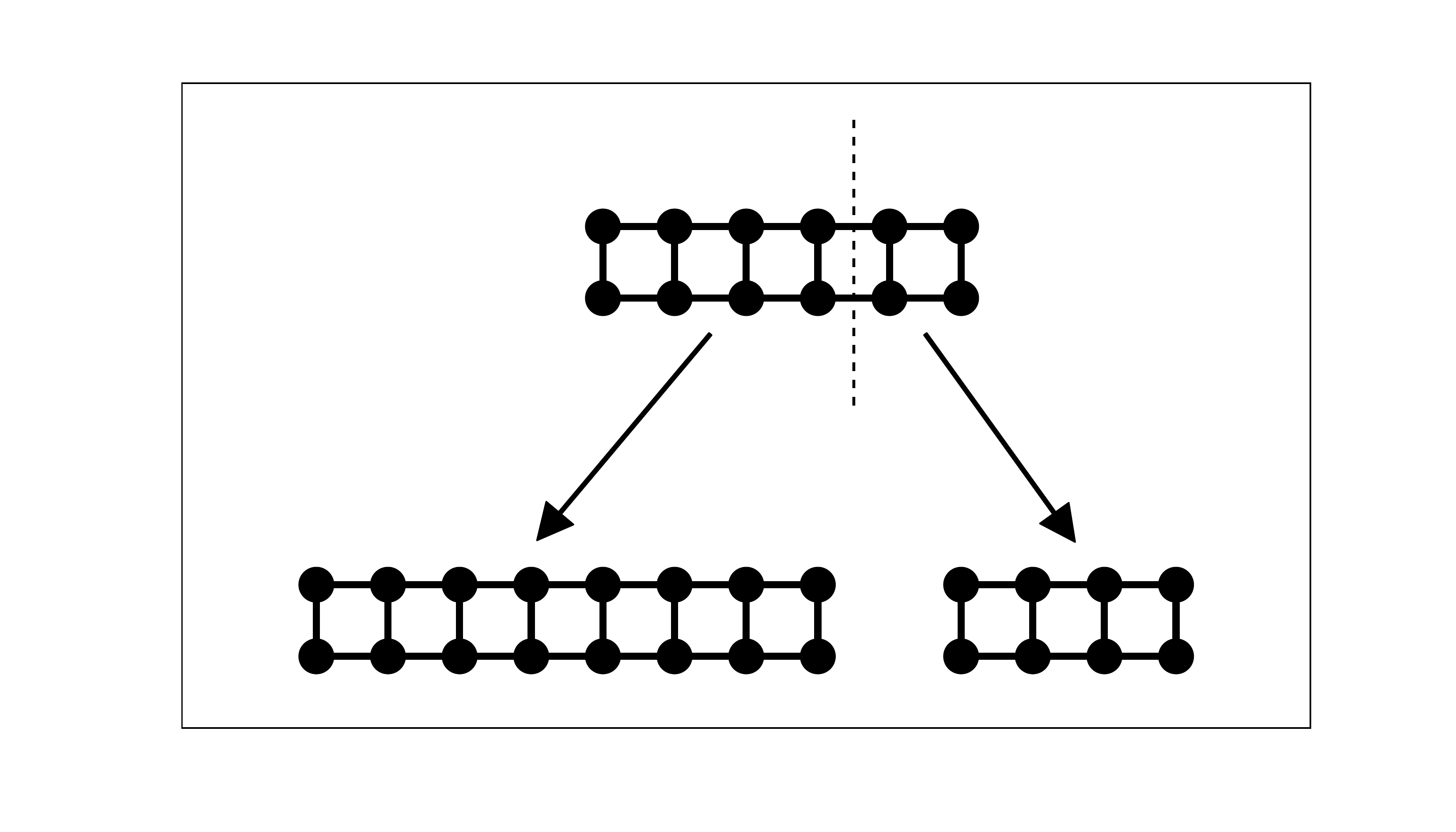}
\caption{Scheme of the open ladder lattice and demonstration of the division. Ladder of the $L=6$ is divided into ladders with $L=4$  and $L=2$ and by doubling the length we are 
creating two ladders of $L=8$ and $L=4$.}\label{rys:SchemeLadder}
\end{figure}

\begin{figure}[!ht]
\center
\includegraphics[width=1.0\textwidth]{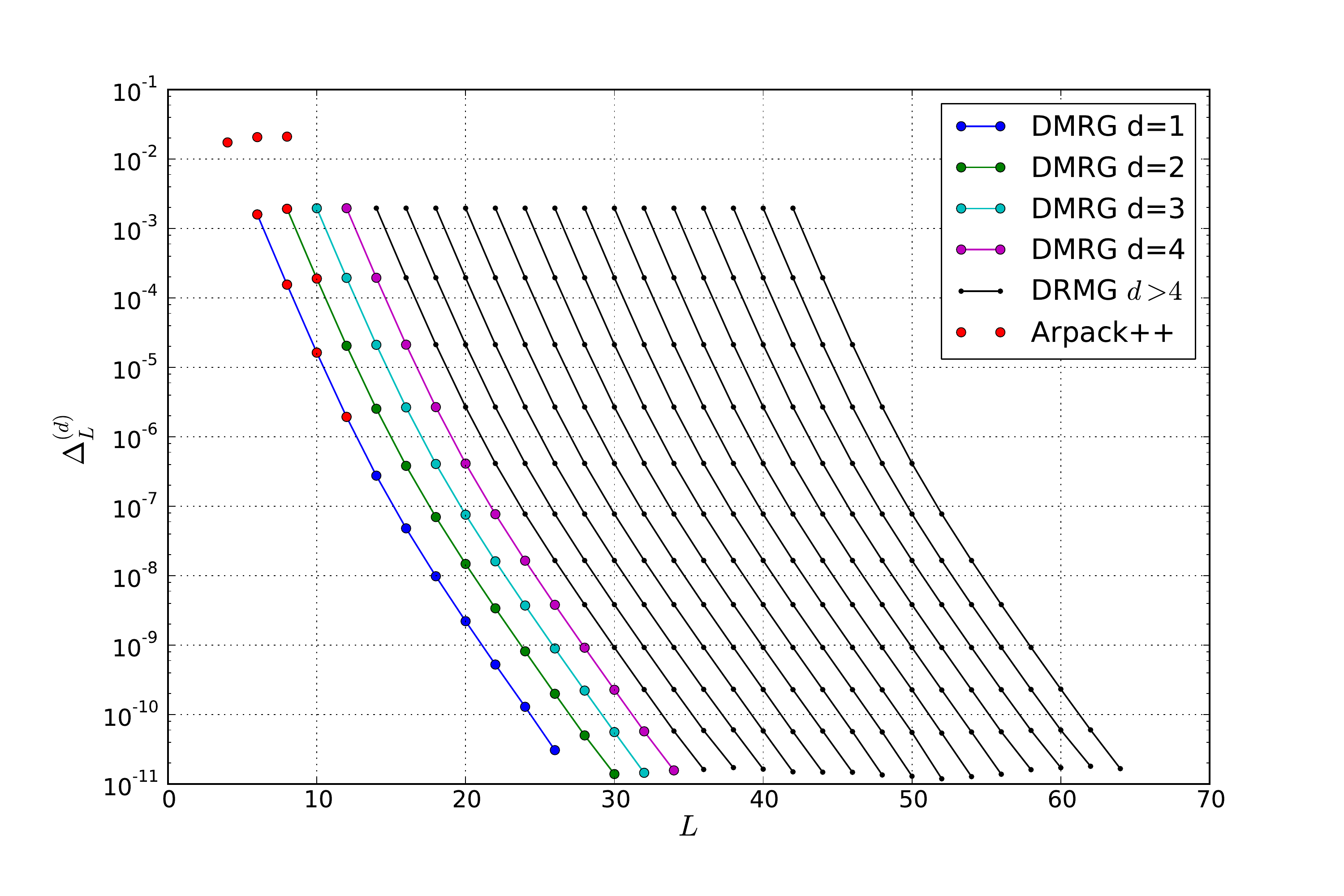}
\caption{ Square ladders: plot of $\Delta^{(d)}_L$ as a function of $L$ for diferent values of $d$. The illustrated data have been obtained
 by exact diagonalization and DMRG.  For not-too-large systems (30-site systems divided near the middle) the difference $\Delta^{(d)}_L$ become smaller than the numerical error of the DMRG algorithm.}\label{rys:res_ladder}
\end{figure}

We examine here the value of the difference $\Delta^{(d)}_L$ defined by~(\ref{LSchainsIII}), 
similarly as we did in the case of the open chain systems. The only difference is that 
now by ladder of length $L$ we refer to square lattice of size $2\times L$ (for details, see Fig.~\ref{rys:SchemeLadder}). The results are presented in Fig.~\ref{rys:res_ladder}.

It is seen that the values of $\Delta^{(d)}_L$ are {\em very small}. 
Typically, they are $10^3$ -- $10^4$ times smaller than
corresponding values for chains of the same size.

 For not-too-large systems (30-site systems divided near the middle) the difference $\Delta^{(d)}_L$ become smaller than the numerical error of the DMRG algorithm. Within the method precission, the difference always was greater than zero.

%%%%%%%%%%%
\newpage
\subsection{Pyrochlore and ladders  with crossing bonds }
We test here another systems to which inequality (\ref{MainIneq}) applies. They are:
ladders with crossing bonds and 'pyrochlore' ladders. Simillarly  to the previous cases of chains and open ladders, data were obtained using exact diagonalization and DMRG method. The ED results are collected in Table~\ref{tab:quasi2d} in Appendix~A.

\subsubsection{Ladders with crossing bonds}
\begin{figure}[ht]
\center
\includegraphics[width=1.0\textwidth]{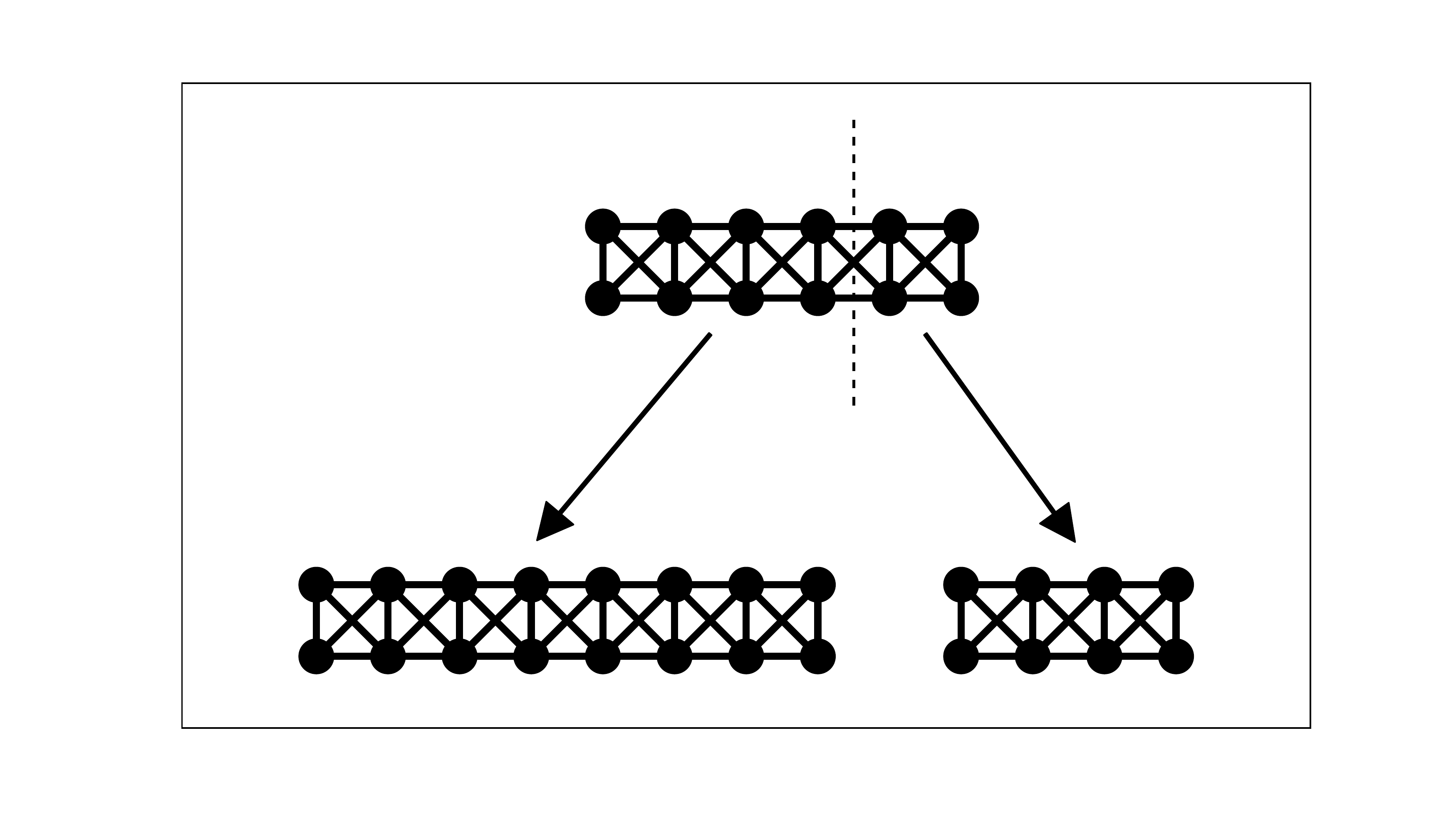}
\caption{Scheme of the open ladder with crossing bonds and demonstration of the division. Ladder of the $L=6$ is divided into ladders with $L=4$  and $L=2$ and by doubling the length we are creating two ladders of $L=8$ and $L=4$.}\label{rys:SchemeXLadder}
\end{figure}

For the ladder with crossing bonds, the horizontal and vertical couplings are 1, whereas diagonal couplings are equal to $\frac{1}{2}$. For details of the system division,
 see Fig.~\ref{rys:SchemeXLadder}.

\begin{figure}[H]
\center
\includegraphics[width=1.0\textwidth]{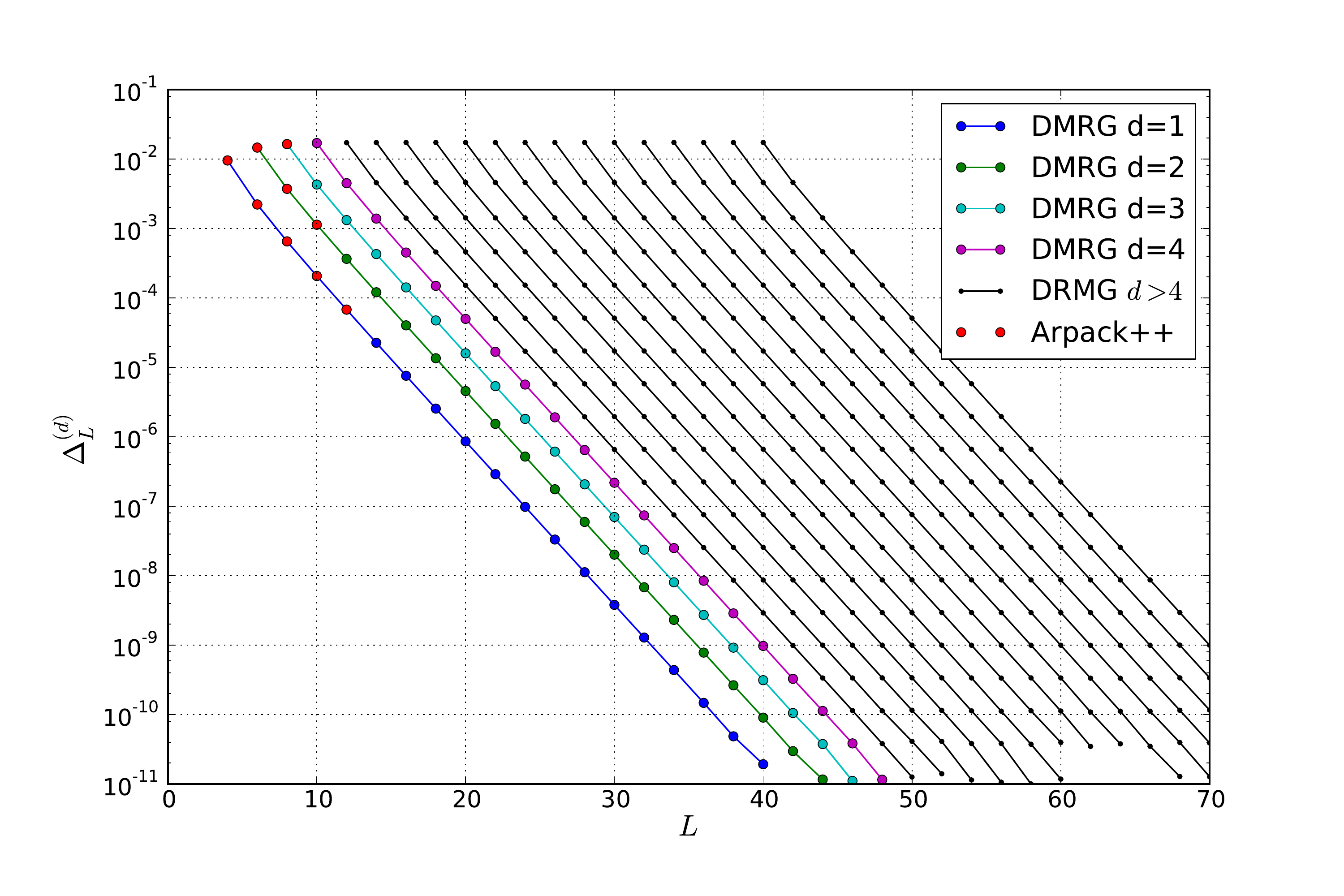}
\caption{Ladders with crossing bonds: plot of $\Delta^{(d)}_L$ as a function of $L$ for diferent values of $d$. The illustrated data have been obtained
 by exact diagonalization and DMRG.  For not-too-large systems (40-site systems divided near the middle) the difference $\Delta^{(d)}_L$ become smaller than the numerical error of the DMRG algorithm.}\label{rys:res_xOpLattice}
\end{figure}

%??? DO SPRAWDZENIA
Numerical results are presented in Fig.~\ref{rys:res_xOpLattice}. Values of $\Delta^{(d)}_L$ -- as a rule -- are larger than for ordinary ladders, but still are $10^4$ -- $10^5$ times smaller than for chains.
% KONIEC DO SPRAWDZENIA

\subsubsection{Pyrochlore ladders}
For pyrochlores, all coupling constants are equal to 1, and 
This was motivated by the fact that for rectangular lattices of larger width, the inequality
(\ref{MainIneq}) applies when diagonal $J_d$ and horizontal\slash{}vertical $J$ couplings
satisfy: $\frac{J_d}{J}\leq\frac{1}{2}$ \cite{Schupp}, \cite{LSW}.

\begin{figure}[ht]
\center
\includegraphics[width=1.0\textwidth]{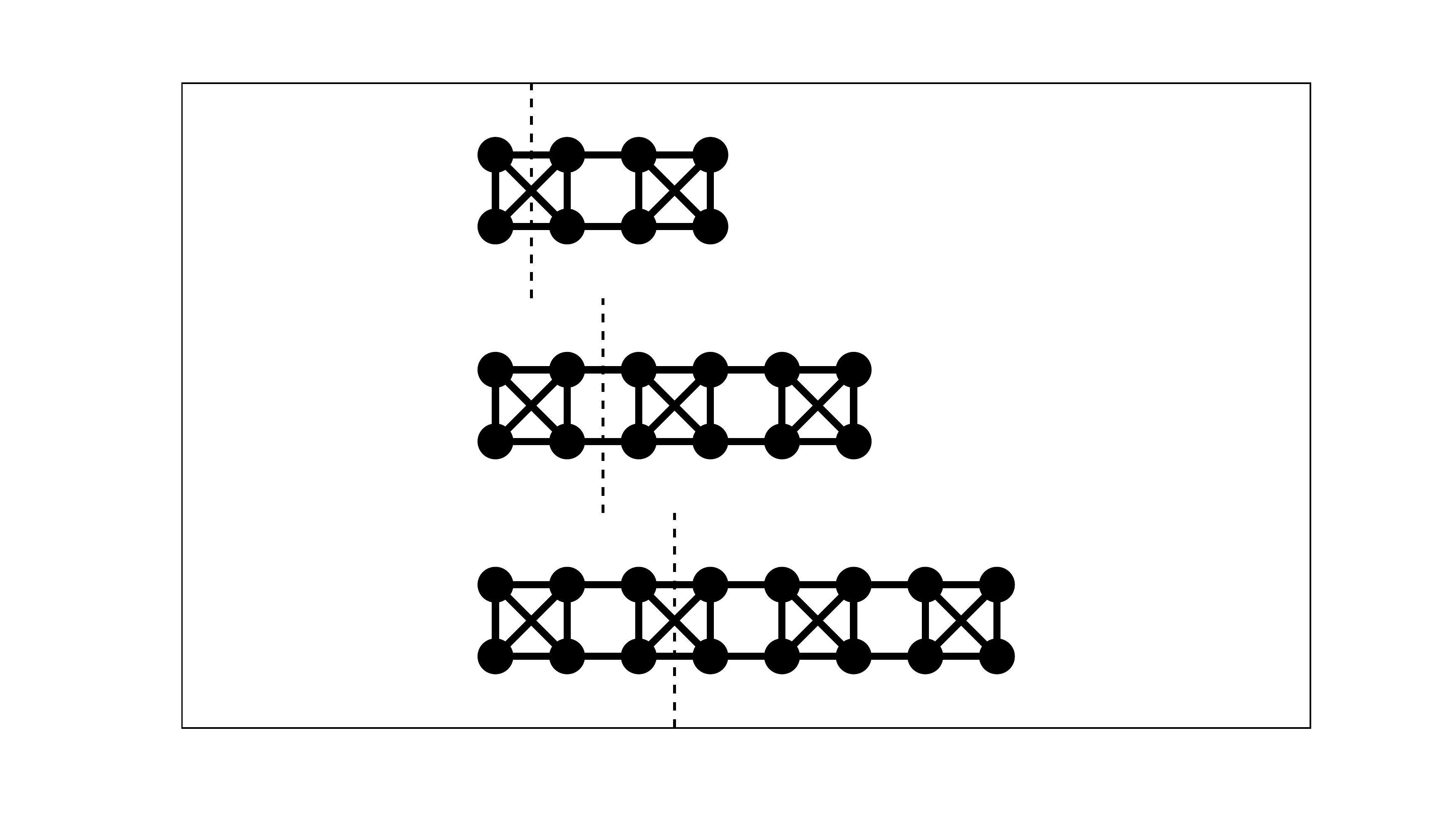}
\caption{Possible slices for pyrochlore lattices. They are discussed in the Appendix B.}\label{rys:SchemePyros}
\end{figure}

Fig.~\ref{rys:SchemePyros} shows a family of system divisions that was chosen. They exhaust all
interesting possibilities; for details, see Appendix B.

\begin{figure}[H]
\center
\includegraphics[width=1.0\textwidth]{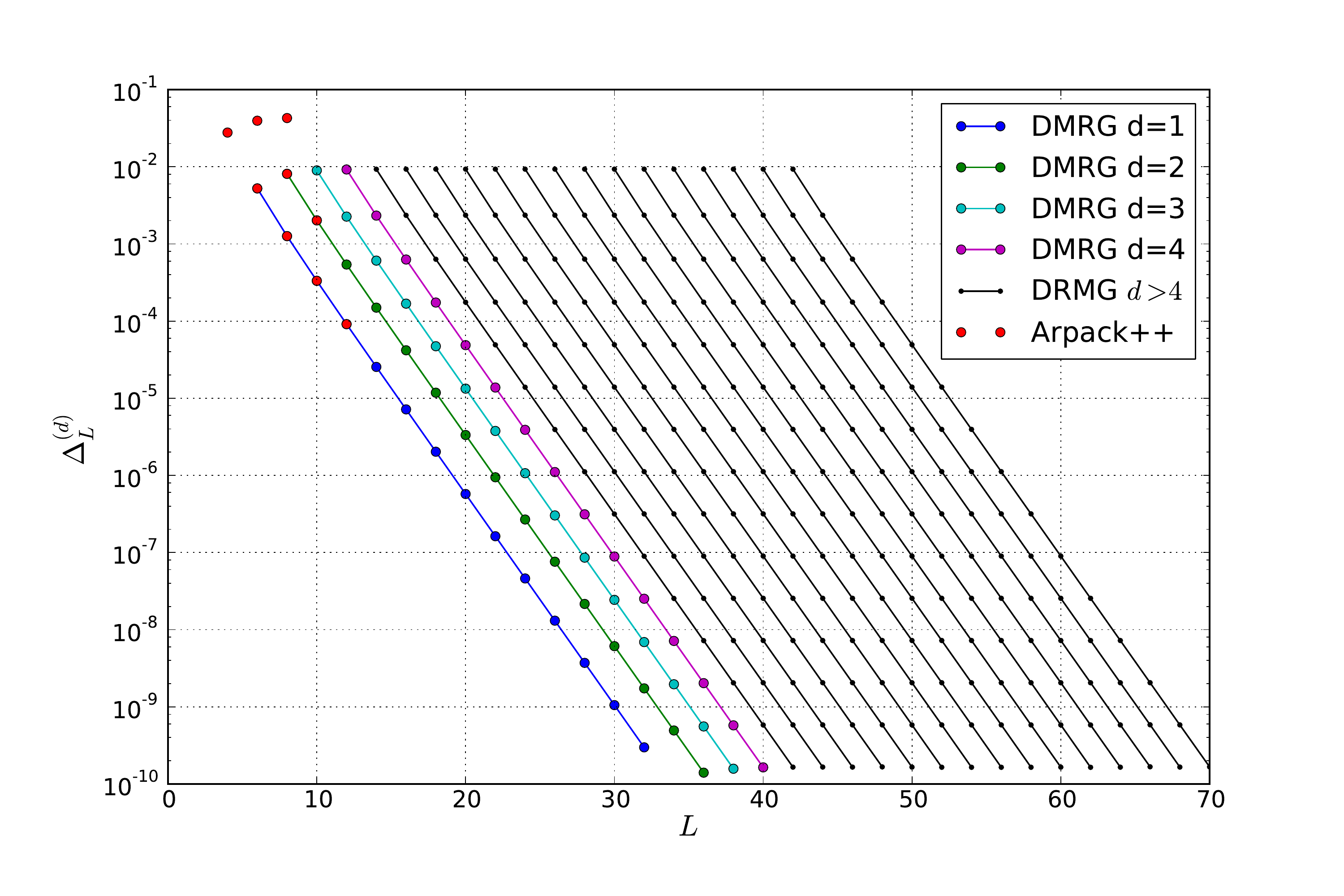}
\caption{Pyrochlore ladders: plot of $\Delta^{(d)}_L$ as a function of $L$ for diferent values of $d$. The illustrated data have been obtained
 by exact diagonalization and DMRG.  For not-too-large systems (40-site systems divided near the middle) the difference $\Delta^{(d)}_L$ become smaller than the numerical error of the DMRG algorithm.}\label{rys:res_pyros}
\end{figure}

Ground-state energies  are collected in Table~\ref{tab:quasi2d}, whereas values of
$\Delta_L$ (defined analogously as for square ladders) are illustrated in Fig~\ref{rys:res_pyros}.

One can ask how the difference $\Delta$ depends of the system size in the case of ladders.
Figures \ref{rys:res_ladder}, \ref{rys:res_xOpLattice} and \ref{rys:res_pyros} suggest that 
 in all cases this dependence is {\em exponential} one.
We have fitted the data to exponential
function:
\be
\Delta \sim \exp(-\alpha L)
\label{DeltaLadders}
\ee
The fit was moderately good for simple ladders (here $\alpha=0.68(5)$) and very good for 
'crossed' ($\alpha=0.541(7)$) and pyrochlore ($\alpha=0.630(5)$) ladders. 

\begin{figure}[H]
\center
\includegraphics[width=1.0\textwidth]{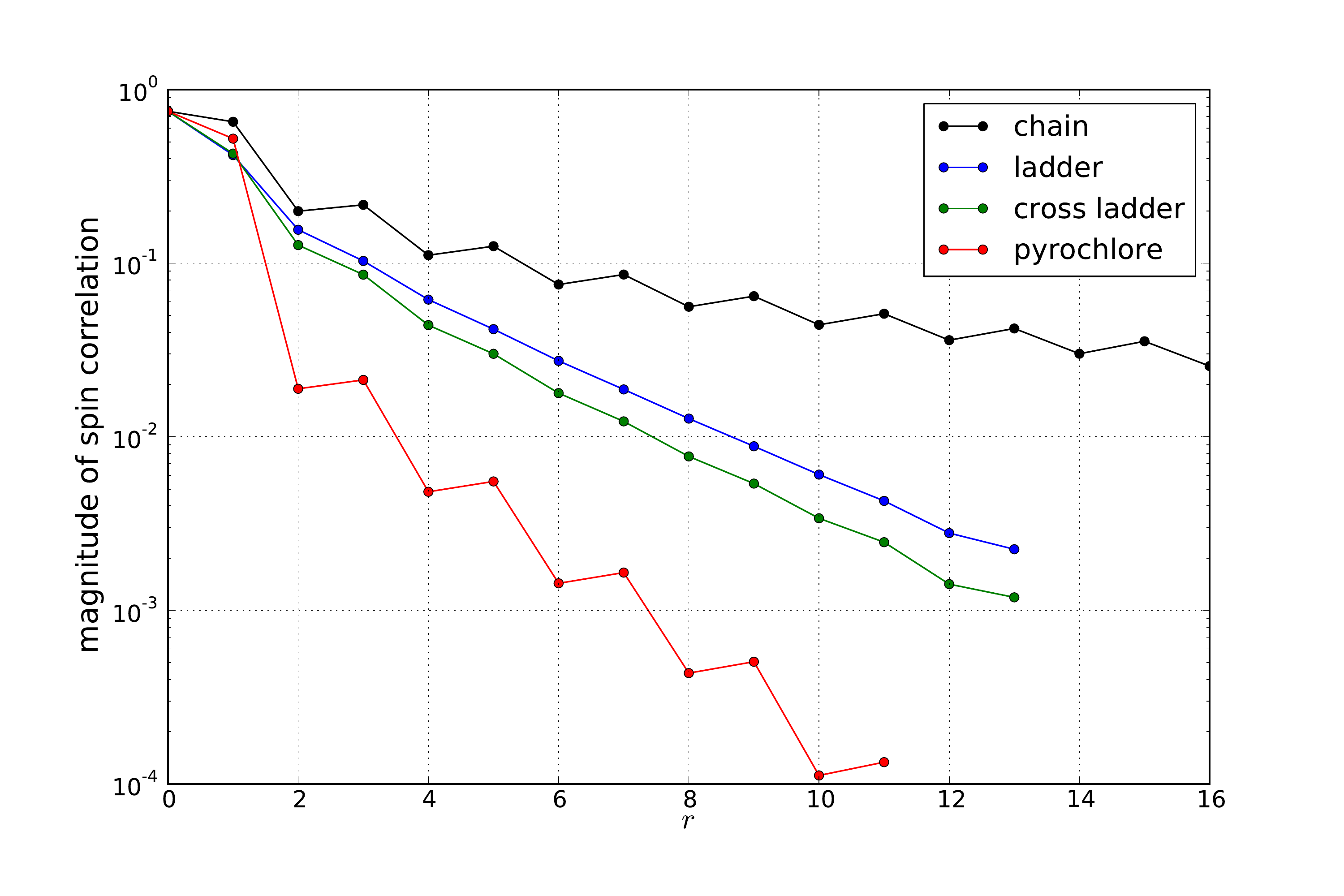}
\caption{Correlation function $\brak \Sgr_1\cdot\Sgr_j\kket$ as a function of inter-site distance $j$.
The correlations tend to zero in {\em algebraic} manner for the chain, whereas for ladders, they
fall-off {\em exponentially}.}
\label{rys:correlations}
\end{figure}

We have observed that for all three sorts of ladders, values of $\Delta^{(d)}_L$ are
much smaller than for chains. Which is the origin of this difference?
One can guess that the more the $L$ and $R$ subsystems are independent (uncorrelated),
the smaller difference $\Delta^{(d)}_L$ is. 
So it is natural to compare the value of $\Delta^{(d)}_L$ with behaviour of 
{\em correlations} in system of consideration. We calculated some (selected) correlation
functions $\brak \Sgr_i\cdot\Sgr_j\kket$ in ladders (see Fig. \ref{rys:correlations}).

We have checked that in all cases
the decay of correlations was {\em exponential} one.

Such a behaviour is consistent with (extension of) Haldane rule
\cite{Haldane}, \cite{Sachdev}, \cite{KolezhukMikeska}. On the other hand, it is well known 
that in the Heisenberg chain the decay of correlations is algebraic. So correlations for chains are
stronger than for ladders, and it turns out that $\Delta^{(d)}_L$ is larger for chains than for ladders.

We made also another observation. Namely,  for some pyrochlore systems, 
the difference $\Delta^{(d)}_L$ is {\em equal to zero} (within numerical precision),
what means that the inequality {\em saturates}! It is a consequence of the fact 
that for pyrochlore systems, the addition of the pair of sites at the end of the ladder
changes the energy by a {\em constant value} (equal to $-0.75$),
 independent of the ladder size 
(see Table~\ref{tab:quasi2d}). In this case, the correlation function of the boundary spin
with all other spins is exactly zero (again within numerical precision).
 It would be very interesting to understand
these facts theoretically.

%%%%%%%%%%%
\subsection{Rectangles of width 3, 4 and 5}

\begin{figure}[H]
\center
\includegraphics[width=1.0\textwidth]{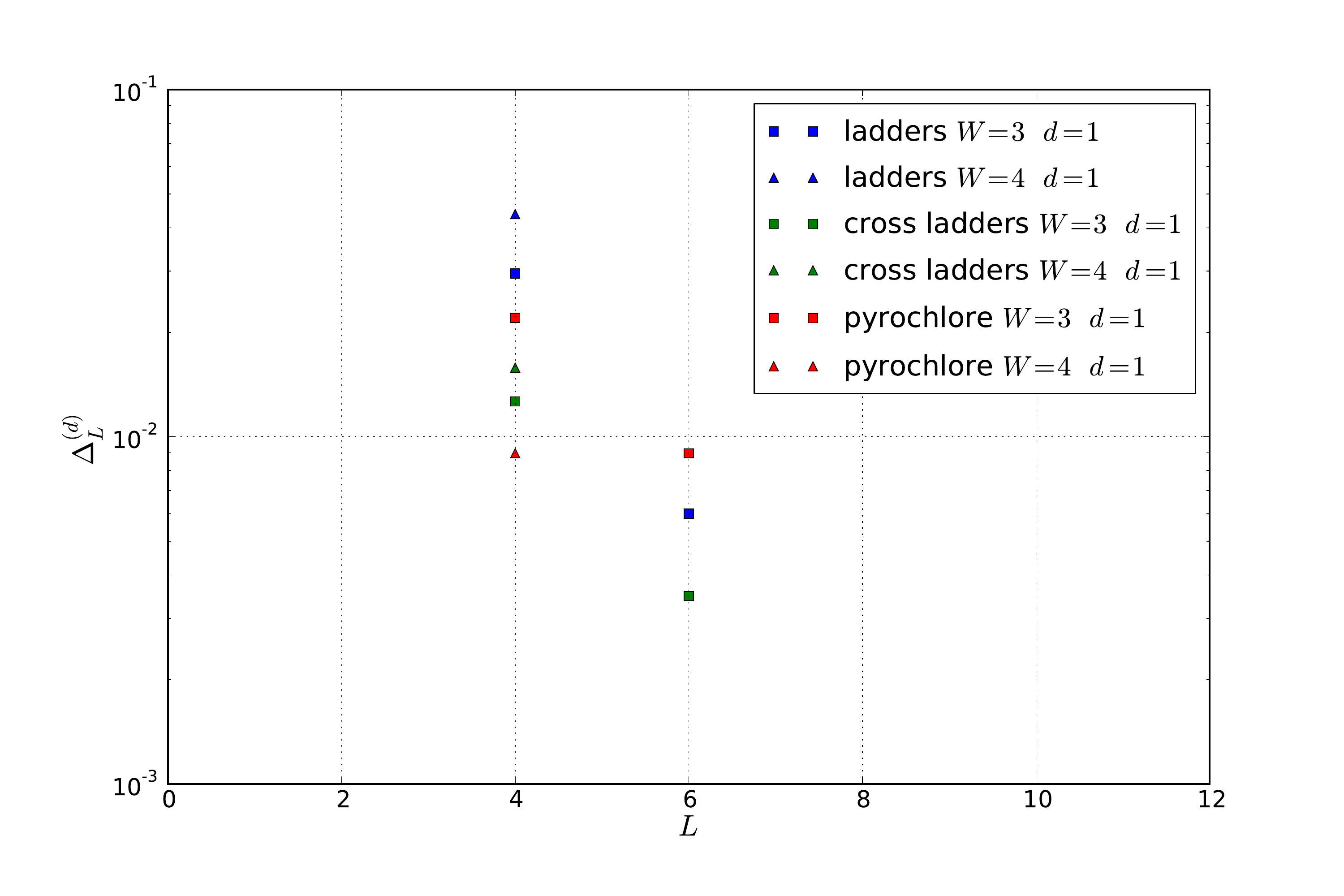}
\caption{Differences of energies for strips of width 3, 4 and 5}
\label{rys:SzerokosciGeq3}
\end{figure}

The results of exact diagonalization for small rectangular subsets of square, pyrochlore
 and  square-with-crossing-bonds lattices are collected in the  Table~\ref{tab:quasi2d}. 
 The differences between two sides of inequality are presented 
in Fig. \ref{rys:SzerokosciGeq3}.
 
It can be seen that differences are larger for square lattice systems than for 
ladders and chains. It can be again related with behaviour of correlation functions, 
which are expected to tend for large distances to non-zero constant \cite{AF2dHeisGS},
\cite{Sandvik}.
 So the correlations
are larger here, and the values of $\Delta^{(d)}_L$ are larger, too.

%KONIEC SPRAWDZENIA

%%%%%%%%%%%
\subsection{Some conjectures}
%%%%%%%%%%%
Let us summarize our observations:

The smallest values of $\Delta^{(d)}_L$ is observed for {\em i)} all ladders and for pyrochlore as well as
X lattice. The larger value of $\Delta^{(d)}_L$ takes place for {\em ii)} chains,
 and the largest one -- for {\em iii)} quasi-2d square lattice.

These results seem to be related with the behaviour of {\em spin correlations} 
in the system considered.
For the {\em i)} case, the spin correlations $\brak  \Sgr_i\cdot\Sgr_j \kket$ tends to zero 
{\em exponentially} with the distance $|i-j|$.
 For {\em ii)}, the fall-off of correlations has power law ,
  whereas for {\em iii)}, correlations are expected to tend to a constant 
 
It seems for us that this inter-relation can be true in the more general situations and we 
formulate the following conjecture.

{\bf Conjecture} {\em rough version}. The faster fall-off of the spin correlations in the system,
 the smaller difference between two sites of the inequality (\ref{MainIneq}).

{\bf Conjecture} {\em more precise version for chains and ladders}.
 If the fall-off of the correlation function is exponential,
then the difference $\Delta^{(d)}_L$ as a function of $L$ behaves in an
analogous manner. If the correlation function decays with
the distance acoording to the power law, then the  difference $\Delta^{(d)}_L$ as a function of $L$
also decreases according to the power law.

%%%%%%%%
\section{Summary}
\label{sec:Summary}
%%%%%%%%
In the paper, the numerical value of Lieb-Schupp inequality has been tested for 
spin systems in various geometries: chains, ladders and quasi-two-dimensional lattices.
The tools used were Exact Diagonalization (for all classes of systems up to 28 spins) and 
DMRG (for ladders and chains).

It has been checked that Lieb-Schupp inequality has been fulfilled in all cases
(it must be so as this inequality is rigorous). But the value of difference between two
sides of inequality $\Del^{(d)}_L$ was strongly dependent of the class of system. The largest value 
was observed for quasi-2d systems, smaller -- for chains, and the smallest -- for
ladders. It has been conjectured that the value of $\Del^{(d)}_L$ is related with the fall-off 
of correlations: The smaller value of $\Delta^{(d)}_L$, the faster fall-off of two-point
correlation functions. In some cases (pyrochlore ladders) the value of $\Del$ was
equal to {\em zero} -- i.e. the inequality {\em saturates}. In these cases, also 
certain correlation functions has been exactly equal to zero.

It would be very interesting to understand these facts theoretically. Moreover, in the
cases where $\Del^{(d)}_L=0$ and correlation functions are equal to zero, perhaps it would be
possible to construct the ground state of the system in the spirit of Matrix Product States
\cite{KolezhukMikeska}, \cite{ZKS1}, \cite{ZKS2}.

%The relation of an analogon of solvation forces for ground state energies with Lieb-Schupp 
%inequalities has also been discussed.

%%%%%%%%%%%%%%%%%%%%%%%%%%%%%%%%%%%%%%%%%%%%%%%%%%%%%%%%%%%%%%%%%%%%%%%%%
\newpage
\section*{Appendix A}
\label{Appendix A}

\begin{table}[!htp]
\centering
\begin{tabular}{|c|r|} \hline
%   \multicolumn{3}{|c|}{łańc. otwarte} \\\hline
$N$ & $E_N$ %& $E_N/N$ 
\\\hline 
 2 &    -0.750000000000 \\
  3 &    -1.000000000000 \\
  4 &    -1.616025403784 \\
  5 &    -1.927886253318 \\
  6 &    -2.493577133888 \\
  7 &    -2.836239680687 \\
  8 &    -3.374932598688 \\
  9 &    -3.736321706379 \\
 10 &    -4.258035207283 \\
 11 &    -4.632093302360 \\
 12 &    -5.142090632841 \\
 13 &    -5.525322097084 \\
 14 &    -6.026724661862 \\
 15 &    -6.416920491794 \\
 16 &    -6.911737145575 \\
 17 &    -7.307408708036 \\
 18 &    -7.797011068537 \\
 19 &    -8.197105741633 \\
 20 &    -8.682473334399 \\
 21 &    -9.086218400935 \\
 22 &    -9.568075875984 \\
 23 &    -9.974886805423 \\
 24 &   -10.453785760410 \\
 25 &   -10.863209352260 \\
 26 &   -11.339579652755 \\
 27 &   -11.751257222131 \\
 28 &   -12.225440548603 \\\hline
\end{tabular}
\caption{Ground state energies in the open chain lattice. Coupling constant $J = 1$ and $N$ represents lengths of the chains.}
\label{tab:gsChains}
\end{table}

\begin{table}[!htp]
\centering
\begin{tabular}{|c|c|r|r|r|r|} \hline
Nx  & Ny & Square & Pyrochlore, case A & Pyrochlore, case B 
& X lattice
\\\hline
2 & 2 &  -2.0000000000000 &  -1.5000000000000 &  -2.0000000000000 &  -1.7500000000000 \\
2 & 3 &  -3.1293852415718 &  -2.7500000000000 &  -2.7500000000000 &  -2.6778862533180 \\
2 & 4 &  -4.2930664566570 &  -3.5000000000000 &  -4.0277505942154 &  -3.6418298745657 \\
2 & 5 &  -5.4467120643352 &  -4.7777505942154 &  -4.7777505942154 &  -4.5873084880937 \\
2 & 6 &  -6.6034724753869 &  -5.5277505942154 &  -6.0607411404916 &  -5.5431961748705 \\
2 & 7 &  -7.7593260611500 &  -6.8107411404916 &  -6.8107411404916 &  -6.4933181096298 \\
2 & 8 &  -8.9154711235558 &  -7.5607411404916 &  -8.0949932311853 &  -7.4467788894730 \\
2 & 9 & -10.0715341613484 &  -8.8449932311853 &  -8.8449932311853 &  -8.3983424061390 \\
2 & 10& -11.2276251173666 &  -9.5949932311853 & -10.1295778777187 &  -9.3510128680377 \\
2 & 11& -12.3837088687482 & -10.8795778777187 & -10.8795778777186 & -10.3030475515486 \\
2 & 12& -13.5397954074066 & -11.6295778777187 & -12.1642537019725 & -11.2554538123736 \\
2 & 13&	-14.6958813681522 & -12.9142537019725 & -12.9142537019725 & -12.2076455365328 \\
2 & 14& -15.8519676317127 & -13.6642537019726 & -14.1989549790545 & -13.1599626306679 \\
\hline		
3 & 3 &  -4.7493272585528 &  -4.0087848535303 &  -4.0087848535303  & -3.9593399973975 \\
3 & 4 &  -6.6916801935149 &  -5.6617068232824 &  -5.6617068232824  & -5.5345034217058	\\
3 & 5 &  -8.3876285183968 &  -6.9910226671666 &  -6.9910226671666  & -6.8685484091210 \\
3 & 6 & -10.2835182238578 &  -8.5954218204916 &  -8.5954218204916  & -8.4037660947387 \\
3 & 7 & -12.0072308867337 &  -9.9647656528603 &  -9.9647656528602  & -9.7629674917248   \\
3 & 8 & -13.8813610992452 & -11.5380986367773 & -11.5380986367773  &-11.2765046416748 \\
\hline
4 & 4 &  -9.1892070651929 &  -7.3280745721674 &  -8.1022525727023  & -7.5055569500810 \\
4 & 5 & -11.6515708351580 &  -9.7410214922860 &  -9.7410214922860  & -9.4307787759889 \\
4 & 6 & -14.1291468644466 & -11.3817745234209 & -12.1857227796133  &-11.3850974919405 \\
5 & 5 & -14.6961464371187 & -12.0391686609399 & -12.0391686609399  &-11.7667640193786\\
\hline

\end{tabular}
\caption{Ground-state energies in  quasi-two-dimensional systems: square, pyrochlore
(two kinds) and X (square-with-crossing-bonds) lattices}
\label{tab:quasi2d}
\end{table}

%%%%%%%%%%%%%%%%%%%%%%%%%%%%%%%%%%%%%%%%%%%%%%%%%%%%%%%%%%%%%%%%%%%%%%%%%
\newpage
\section*{Appendix B}
\label{Appendix B}

The analysis of the Lieb-Schupp inequality in the case of the pyrochlore ladder system may seem
 to be a complicated task. First of all there exist two types of system -- one that begin with the crossing bonds (denoted as $A$) and one that begin with simple bonds (denoted as $B$). Secondly the line of division initial system may go through a crossing bonds (like in the ladder with the crossing bonds) or through only vertical bonds (like in a simple square ladder case). 

Fortunately analysing data more closely and taking into account behaviour
 of a correlation functions allows us to choose only one family of possible divisions, which
  is equivalent to all others.

The goal of this appendix is to present an argumentation based of the empirical observations
 that the choice of this particular family of divisions is justified. 

\paragraph{Observation 1} On the beginning let take a look on the systems of the even length. We will denote this systems with $A_s$ and $B_s$, because they have a reflection symmetry -- if the system beging with a crossing bonds (type $A$) it also end with a crossing bonds or if the system begin with a simple vertical bond (type $B$) it also end the same way.

\begin{figure}[!htp]
\center
\includegraphics[width=1.0\textwidth]{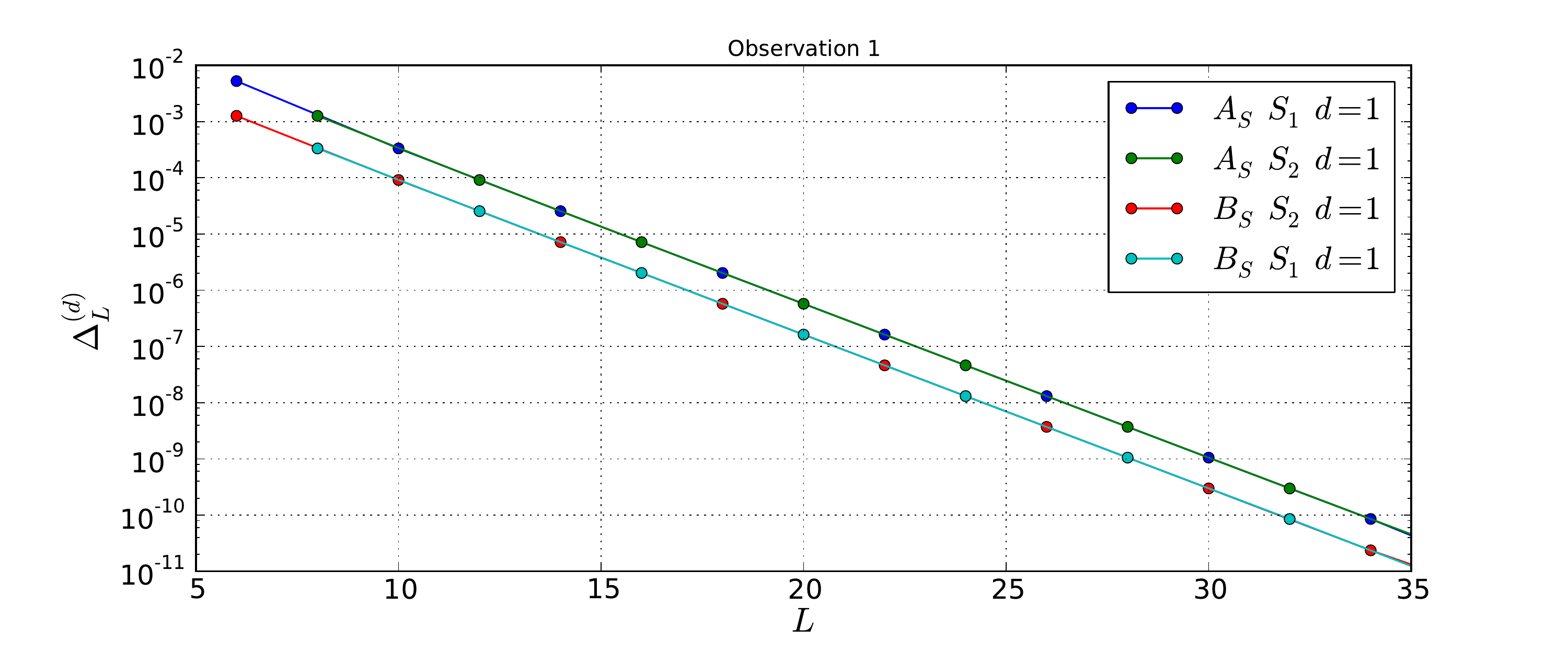}
\caption{Illustration of the observation 1. Different types of slice lies on the same line which allows them to by analysed as a one family of divisions.}
\label{rys:observation1}
\end{figure}

Furthermore let denote by $S_1$ a slice that goes through a simple vertical bonds and by $S_2$ a slice that goes through a crossing bonds. Making plot of a value of $\Delta_L^{(d)}$ for this families [Fig.\ref{rys:observation1}] we notice that slices $S_1$ and $S_2$ can be treated during analyse as one family.

For an example in Fig.~\ref{rys:SchemePyros} we present system of the type $A_s$ with slices of the type $S_1$ in the top and bottom situations and of the type $S_2$ in the middle situation.

\paragraph{Observation 2} System of an odd length $A$ is identical as a system $B$. Let denote this system as $A_a$ to make is distinguished from $A_s$ and $B_s$. From the data in Table~\ref{tab:quasi2d} we can observe that adding two sites on the end of a system $A_a$ (for example coming from $2\times 3$ to $2\times 4$ system or~from~$2\times 5$~to~$2\times 6$ system) the ground state energy change~by~a~constant~value~$-0.75$. It is also worth to notice that adding two spins on the edge of the $B_s$ type system also change ground state energy only by the same constant. 

This observation is consistent with the fact that correlation of the edge spins in the type $A_s$ systems is zero with every other spin in the system\footnote{But not between the two spins on the same edge.}.

It is very easily to proof that if ground state energy of the system that ends with simple bond differs only by a constant from the situation, that this edge is extented with two spins to form a crossing bond, the value of $\Delta_L^{(d)}$ do not change. 

The illustration on this may be seen in Fig.~\ref{rys:observation1} where the $B_s$ system can be extended to a system $A_s$ by adding two sites on the both system edges.

%%%%%%%%%%%%%%%%%%%%%%%%%%%%%%%%%%%%%%%%%%%%%%%%%%%%%%%%%%%%%%%%%%%%%%%%%
 
%
\end{document}